\documentclass[aps,pra,twocolumn,showpacs,superscriptaddress ]{revtex4-1}
\usepackage{enumerate}
\usepackage{graphicx}
\usepackage{amsmath}
\usepackage{amsfonts}
\usepackage{amssymb}
\usepackage{epstopdf}
\newcommand{\ket}[1]{\left|#1\right\rangle}
\newcommand{\bra}[1]{\left\langle#1\right|}

\newcommand{\be}{\begin{equation}}
\newcommand{\ee}{\end{equation}}
\newcommand{\bea}{\begin{eqnarray}}
\newcommand{\eea}{\end{eqnarray}}

\begin{document}
\today
\title{Realization of adiabatic Aharonov-Bohm scattering with neutrons}
\author{Erik Sj\"oqvist$^{\ast}$}
\email{e-mail: erik.sjoqvist@kemi.uu.se} 
\affiliation{Department of Quantum Chemistry, Uppsala University, Box 518, 
Se-751 20 Uppsala, Sweden}
\affiliation{Department of Physics and Astronomy, Uppsala University, Box 516, 
Se-751 20 Uppsala, Sweden}
\author{Martin Almquist}
\affiliation{Department of Information Technology, Uppsala University, L\"agerhyddsv\"agen 2, 
Se-752 37 Uppsala, Sweden}
\author{Ken Mattsson}
\affiliation{Department of Information Technology, Uppsala University, L\"agerhyddsv\"agen 2, 
Se-752 37 Uppsala, Sweden}
\author{Zeynep Nilhan G\"urkan}
\affiliation{Department of Industrial Engineering, Gediz University, Seyrek, 35665, 
Menemen-Izmir, Turkey}
\author{Bj\"orn Hessmo}
\affiliation{Centre for Quantum Technologies, National University of Singapore,
3 Science Drive 2, 117543 Singapore, Singapore}
\affiliation{Department of Physics, National University of Singapore,
3 Science Drive 2, 117543 Singapore, Singapore}
\begin{abstract}
The adiabatic Aharonov-Bohm (AB) effect is a manifestation of the Berry phase acquired 
when some slow variables take a planar spin around a loop. While the effect has been 
observed in molecular spectroscopy, direct measurement of the topological phase shift 
in a scattering experiment has been elusive in the past. Here, we demonstrate an 
adiabatic AB effect by explicit simulation of the dynamics of unpolarized very slow 
neutrons that scatter on a long straight current-carrying wire.  
\end{abstract}
\pacs{03.65.Vf, 03.75.Be}
\maketitle
\section{Introduction} 
Aharonov and Bohm \cite{aharonov59} pointed out that a charged quantum particle may 
acquire an observable phase shift by circling around a completely shielded magnetic flux. 
This remarkable effect is purely non-classical as the electromagnetic field vanishes at 
the location of the particle, which thereby experiences no Lorentz force. The origin of 
the phase shift is topological: it only depends on the winding number of the particle's 
path around the magnetic flux. 

The original Aharonov-Bohm (AB) setup for electric charge belongs to a larger class 
of topological phase effects. This includes topological phase shifts arising for electrically 
neutral particles in certain electromagnetic configurations \cite{aharonov84,wei95} as 
well as in general quantum systems that undergo adiabatic evolution \cite{berry84}. A 
paradigmatic example of the latter is a molecule that acquires an AB phase shift when 
it reshapes slowly around a conical intersection in nuclear configuration space \cite{mead80}. 
This adiabatic AB effect is imprinted in the spectral properties related to the pseudorotational 
molecular motion, and has been observed \cite{vonbusch98} in the metallic trimer Na$_3$. 
It has further been predicted \cite{lepitit90} in scattering-type chemical reactions, such as 
in the hydrogen exchange reaction $\textrm{H} + \textrm{H}_2$. Due to subtle 
cancellation effects, however, direct observation of the AB effect in molecular 
scattering has been elusive in the past \cite{kendrick00,juanes-marcos05,jankunas13}. 

Matter waves in spatially varying electromagnetic fields is a tool to engineer a 
wide range of quantum effects \cite{zhu06,juzeliunas08,vaishnav08}. If the variation 
of these fields is sufficiently slow, the particle motion is governed by adiabatic gauge 
fields \cite{ruseckas05,lin09,dalibard08} similar to those in molecules and has been 
proposed to give rise to AB phase shifts under certain conditions \cite{huo14,larson09}. 
Here, we develop an adiabatic AB effect for matter waves that 
scatter on a static inhomogeneous magnetic field produced by a current-carrying wire. 
The setup uses the quantum properties of neutrons to induce the AB effect. Earlier 
experimental demonstrations of other topological phase effects, such as the 
Aharonov-Casher \cite{cimmino89} and scalar AB phase shifts \cite{allman92}, as 
well as related Berry phase induced spinor rotations \cite{bitter87,richardson88}, have 
shown that neutrons are suitable to realize AB effects due to their robust internal structure, 
the ease of detecting them with almost 100 \% efficiency, and their weak coupling to the 
environment.  Our setup is similar to the one proposed in Ref.~\cite{frustaglia01} for 
a nanoelectronic system. 

\section{Physical setup}
Imagine a colimated beam of neutrons that scatter on a long straight wire of radius $R$, 
carrying an electrical current $I_w$. The resulting magnetic field is given by Biot-Savart's law 
\begin{eqnarray}
{\bf B} = \frac{\mu_0 I_w}{2\pi R} f(r) {\bf e}_{\theta} , \ \  
f(r) = \left\{ \begin{array}{ll} \frac{1}{r} , & r \geq 1 , \\ 
r , & r < 1 , 
\end{array}
\right. 
\label{eq:bfield}
\end{eqnarray}
where we have assumed that the wire points along the $z$-axis and carries 
a uniform current density. Here, $\theta$ is the polar angle in cylindrical coordinates 
with corresponding basis vector ${\bf e}_{\theta}$, $r$ is the distance from the wire 
in units of $R$, and $\mu_0$ is permeability of vacuum. The magnetic field induces 
a local energy splitting of the neutron spin states, as described by the Zeeman 
Hamiltonian $\mathcal{H} = - \boldsymbol{\mu} \cdot {\bf B} = - \mu \frac{1}{2} 
\boldsymbol{\sigma} \cdot {\bf B}$ with $\mu = - 9.65 \cdot 10^{-27}$ J/T the 
neutron magnetic moment and  $\boldsymbol{\sigma}=\left( \sigma_x,\sigma_y,\sigma_z 
\right)$ the standard Pauli operators representing the neutron spin. The eigenvalues of the 
Hamiltonian are $\pm \frac{|\mu| \mu_0 I_w}{4\pi R} f(r) \equiv \pm V_0 f(r)$, 
which introduce a potential energy barrier (well) of height $V_0$ (depth $-V_0$) at the 
surface of the wire.  

The Hamiltonian relevant for the AB effect is given by the kinetic energy of the neutron in 
the $xy$ plane, plus the Hamiltonian describing the neutron spin, i.e., 
\begin{eqnarray}
H_{\textrm{tot}} = -\frac{\hbar^2}{2m} \nabla_{\bf r}^2 + \mathcal{H} .  
\end{eqnarray}
We write the total wave function as $\ket{\psi ({\bf r},s)} = \varphi_{+} ({\bf r},s) \ket{\chi_{+} 
(\theta)} + \varphi_{-} ({\bf r},s) \ket{\chi_{-} (\theta)}$, where ${\bf r} = (x,y) = r(\cos\theta ,
\sin\theta)$ and $\ket{\chi_{\pm} (\theta)} = \frac{1}{\sqrt{2}} \big( \mp i \ket{\uparrow} + 
e^{i\theta} \ket{\downarrow} \big)$ are the local spin eigenvectors with $\sigma_z 
\ket{\uparrow} = \ket{\uparrow}$, $\sigma_z \ket{\downarrow} = - \ket{\downarrow}$. 
The two spin eigenstates are associated with the potential energies $\pm V_0 f(r)$. Here, 
$s=\hbar t/(2mR^2)$ (dimensionless time) with $m=1.67 \cdot 10^{-27}$ kg being the 
neutron mass. 

To determine the spatial wave functions $\varphi_{\pm} ({\bf r},s)$ we eliminate the 
spin degree of freedom by multiplying the Schr\"odinger equation with $\bra{\chi_{\pm}}$ 
yielding the coupled equations 
\begin{eqnarray}
 & & i \partial_s \varphi_{\pm} ({\bf r},s) = \left[  - \partial_{r}^2 -  
\frac{1}{r} \partial_{r} \right. 
\nonumber \\ 
 & & \left. - \frac{1}{r^2} \left( \partial_{\theta} + \frac{i}{2} \right)^2 
\pm \kappa f(r) + \frac{1}{4r^2} \right] \varphi_{\pm}  ({\bf r},s) 
\nonumber \\ 
 & & + \left( \frac{1}{2r^2} + i \frac{1}{r^2} \partial_{\theta} \right) 
\varphi_{\mp}  ({\bf r},s) ,  
\label{eq:exact}
\end{eqnarray} 
where $\kappa = 2mR^2 V_0/\hbar^2$. 

\begin{figure*}[ht]
\centering
\includegraphics[width=0.39\textwidth]{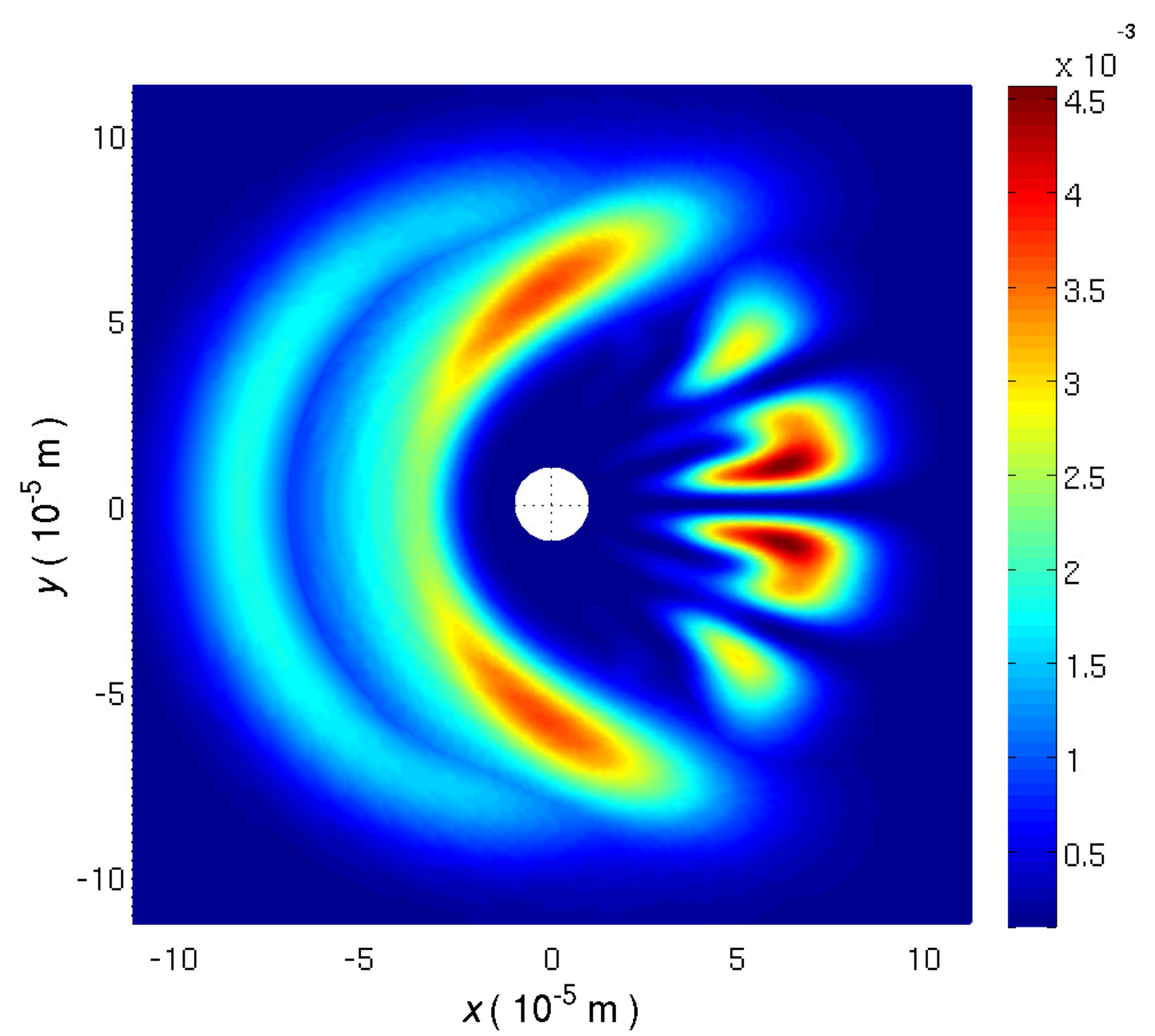}
\includegraphics[width=0.39\textwidth]{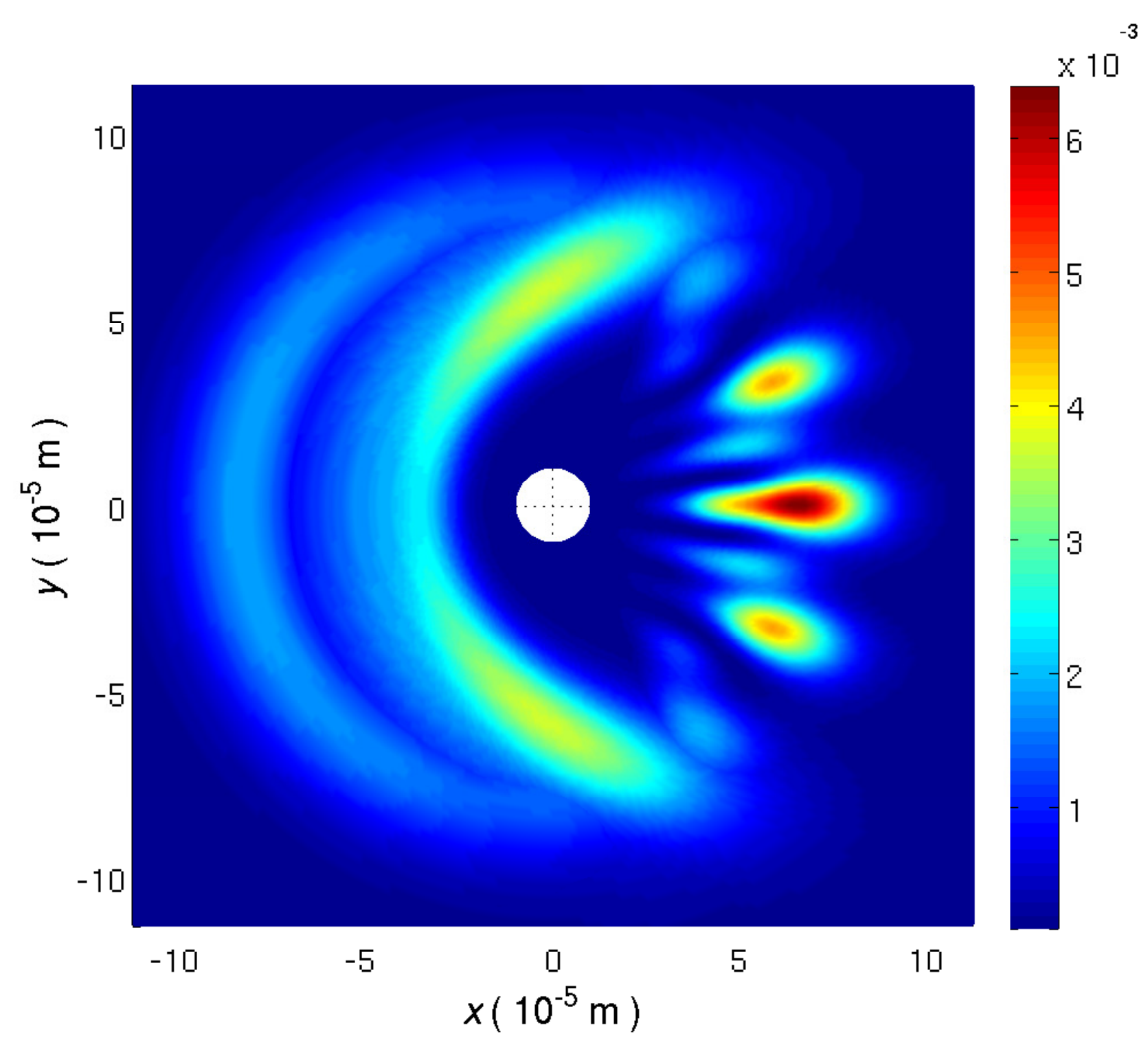}
\caption{Scattered density of unpolarized neutrons moving with incoming velocity $0.02$ 
m/s perpendicular to a very long straight current-carrying wire. Left panel includes the adiabatic 
AB effect. Right panel neglects the adiabatic AB effect. The key difference is the AB-induced 
nodal line in the forward direction $(x>0,y=0)$ that is absent when neglecting the AB effect. 
The wire represented by the white region around the origin has radius $10^{-5}$ m and 
carries $10$ mA electrical current.}
\label{fig:1}
\end{figure*}

The adiabatic regime is characterized by negligible Born-Huang potential $\frac{1}{4r^2}$ 
and off-diagonal coupling terms $\left( \frac{1}{2r^2} \pm i \frac{1}{r^2} \partial_{\theta} 
\right) \varphi_{\mp}  ({\bf r},s)$. For low neutron velocities, the wire is impenetrable and 
the validity of the adiabatic approximation can be checked by comparing the size of the 
scalar potential $\kappa f(r)$ and $1/r^2$ at the surface of the wire. This yields the 
adiabatic condition $\kappa \gg 1$. In this regime, it follows from Eq. (\ref{eq:exact}) 
that the spatial wave packets $\varphi_{\pm} ({\bf r},s)$ are separately determined by the 
effective Schr\"odinger equations 
\begin{eqnarray}
 & & i \partial_s \varphi_{\pm} ({\bf r},s) = \left[  - \partial_{r}^2 -  
\frac{1}{r} \partial_{r} \right. 
\nonumber \\ 
 & & \left. - \frac{1}{r^2} \left( \partial_{\theta} + i\alpha \right)^2 
\pm \kappa f(r) \right] \varphi_{\pm}  ({\bf r},s) 
\label{eq:se}
\end{eqnarray} 
where $\alpha/r = \frac{1}{2r}$ is the Mead-Berry vector potential pointing in the 
$\theta$ direction. 
 
The shift $\partial_{\theta} \rightarrow \partial_{\theta} + i\alpha$ is the key origin  
of the AB effect. It originates from the adiabatic elimination of the spin, which introduces 
the Mead-Berry vector potential \cite{mead80,berry84} ${\bf A} = i \bra{\chi_{\pm}} 
\nabla_{\bf r} \ket{\chi_{\pm}} = -(\alpha /r) {\bf e}_{\theta}$ in the kinetic energy of 
the neutron. This defines the effective magnetic flux 
\begin{eqnarray}
\Phi = \oint_C {\bf A} \cdot d{\bf x} = -2\pi \alpha = -\pi 
\end{eqnarray}
picked up by the phase when the neutron moves around the wire along a loop 
$C$. $\Phi$ vanishes if the neutron does not encircle the wire. The vector potential 
thus corresponds to an AB flux line carrying half a flux unit (semi-fluxon) sitting 
at the conical intersection point at $r=0$.  The flux-induced phase is a topological 
property as it depends neither on the detailed shape of the loop $C$ nor on the 
dynamical parameters of the system, such as the electrical current $I_w$ and the 
speed of the neutron. A consequence of the topological nature of the phase shift is 
that for a non-uniform electrical current density, we can predict that there must be 
an odd number of points at which the magnetic field vanishes inside the wire 
as each conical intersection carries half a flux unit \cite{zwanziger87,aharonov94}. 

\section{Results}
We numerically simulate an unpolarized neutron that scatters on the wire. We solve 
Eq. (\ref{eq:se}) for the two adiabatic channels and calculate the density profile 
\begin{eqnarray}
\rho ({\bf r},s) = \frac{1}{2} \left( \left| \varphi_+ ({\bf r},s) \right|^2 + 
\left| \varphi_- ({\bf r},s) \right|^2 \right)
\end{eqnarray} 
obtained by tracing over the spin. We choose $I_w = 10$ mA and $R = 10^{-5}$ m, 
which yields $\kappa = 29$. The spatial degree of freedom of the neutron is modeled 
as an incoming Gaussian wave packet of $\sim 20$ $\mu$m width, whose centre is 
moving in the $xy$ plane straight on to the wire at relative velocity $v_{xy}$. The wire 
is assumed to be made of an electrically conducting material with a positive neutron 
optical potential that can reflect slow neutrons. An appropriate choice would be copper 
with a positive optical potential corresponding to a critical neutron velocity of about 5.7 m/s. 
We assume neutron velocities $v_{xy}$ well below this critical value, which means that 
we can take $\varphi_{\pm} ({\bf r},s)$ to vanish at $r=1$. With the chosen $\kappa$, 
we expect the adiabatic treatment to be valid, which has indeed been  confirmed numerically 
to a high degree of accuracy by comparing the wave packet dynamics arising from Eqs. 
(\ref{eq:se}) and (\ref{eq:exact}). 

The summation-by-parts--simultaneous approximation term (SBP-SAT) 
method \cite{svard14} is a time-stable well-proven high-order difference methodology 
suitable for solving wave-dominated phenomena \cite{almquist14}. In the present study a 
sixth-order accurate SBP-SAT approximation has been employed, solving Eq. (\ref{eq:se}) 
with high fidelity. 

\begin{figure*}[ht]
\centering
\includegraphics[width=0.39\textwidth]{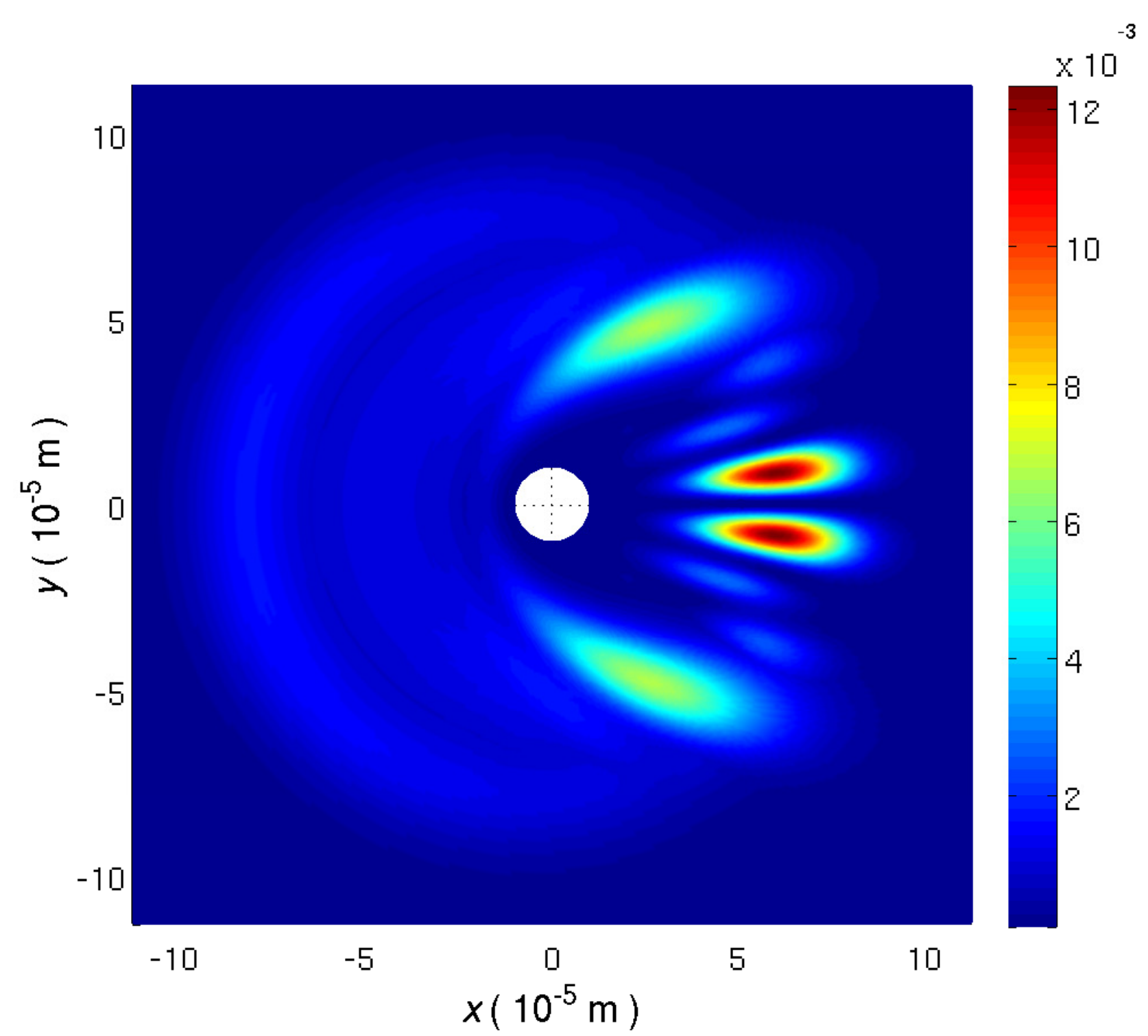}
\includegraphics[width=0.39\textwidth]{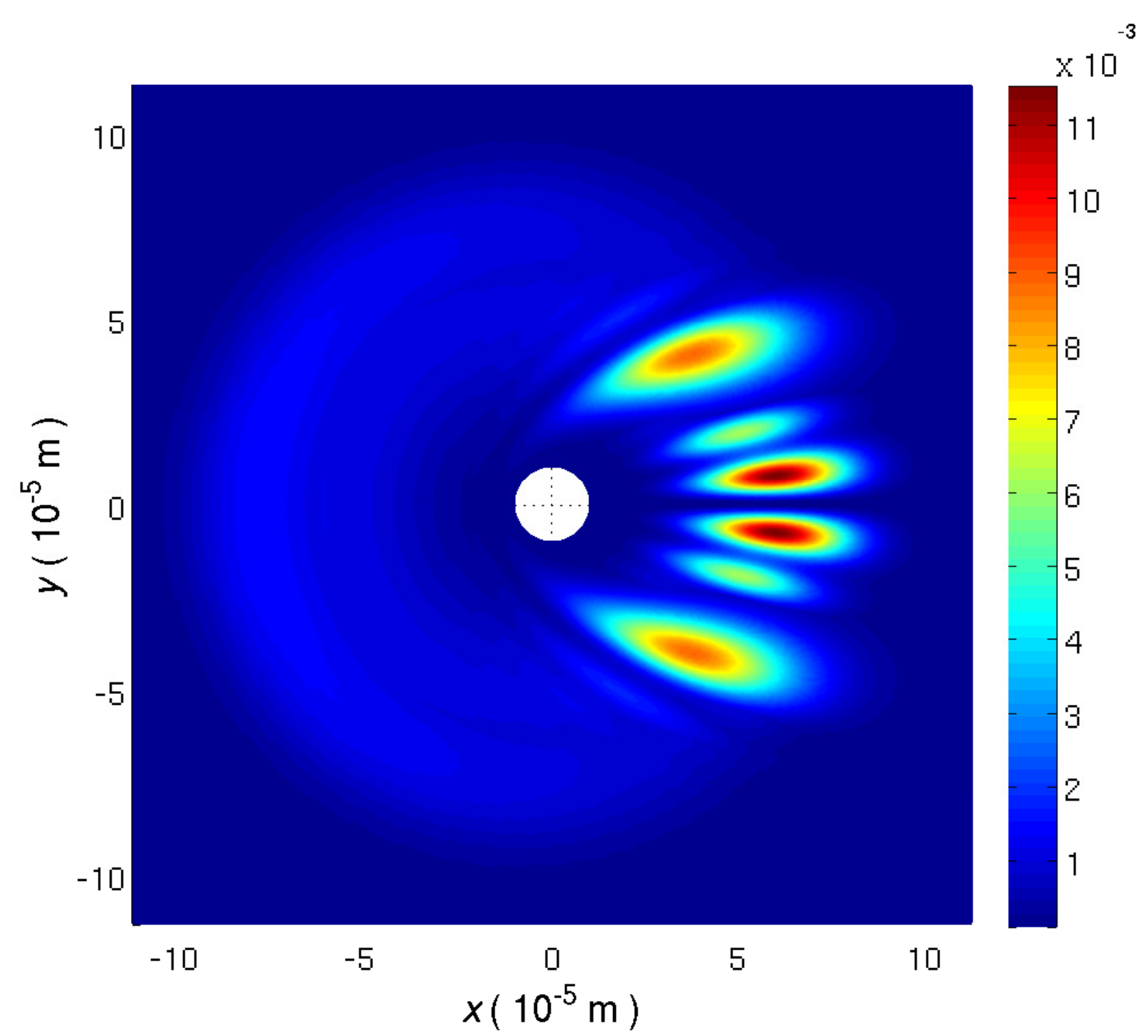}
\caption{Scattered density of unpolarized neutrons at incoming perpendicular incoming 
velocities $0.03$ m/s (left panel) and $0.04$ m/s (right panel) with the same wire radius 
and electrical current as in Fig \ref{fig:1}. The nodal line is clearly visible demonstrating 
the nondispersive nature of the AB effect.}
\label{fig:2}
\end{figure*}

To study the adiabatic AB effect we first compare the dynamics with $v_{xy}=0.02$ m/s for 
$\alpha = \frac{1}{2}$ and $\alpha = 0$, where the latter case corresponds to ignoring 
the adiabatic flux. As seen in Fig. \ref{fig:1} (see also Supplementary Movies 1 and 2), the 
effective AB flux introduces a nodal line in the forward direction, which is not visible when 
$\alpha = 0$. This is the key signature of the adiabatic AB effect related to the effective 
flux line carrying half a flux unit associated with destructive interference in the forward 
direction \cite{huo14,larson09,schon94}. It would be directly detectable by measuring 
the neutron density profile in the $xy$ plane behind the wire. We further perform 
simulations at $v_{xy} = 0.03$ and $0.04$ m/s, shown in Fig. \ref{fig:2}. The nodal line 
persists, which demonstrates the nondispersive nature of the AB effect and opens up for the 
possibility to perform the test even if we allow for a distribution of incoming 
velocities within the adiabatic regime. 

\begin{figure*}[ht]
\centering
\includegraphics[width=0.39\textwidth]{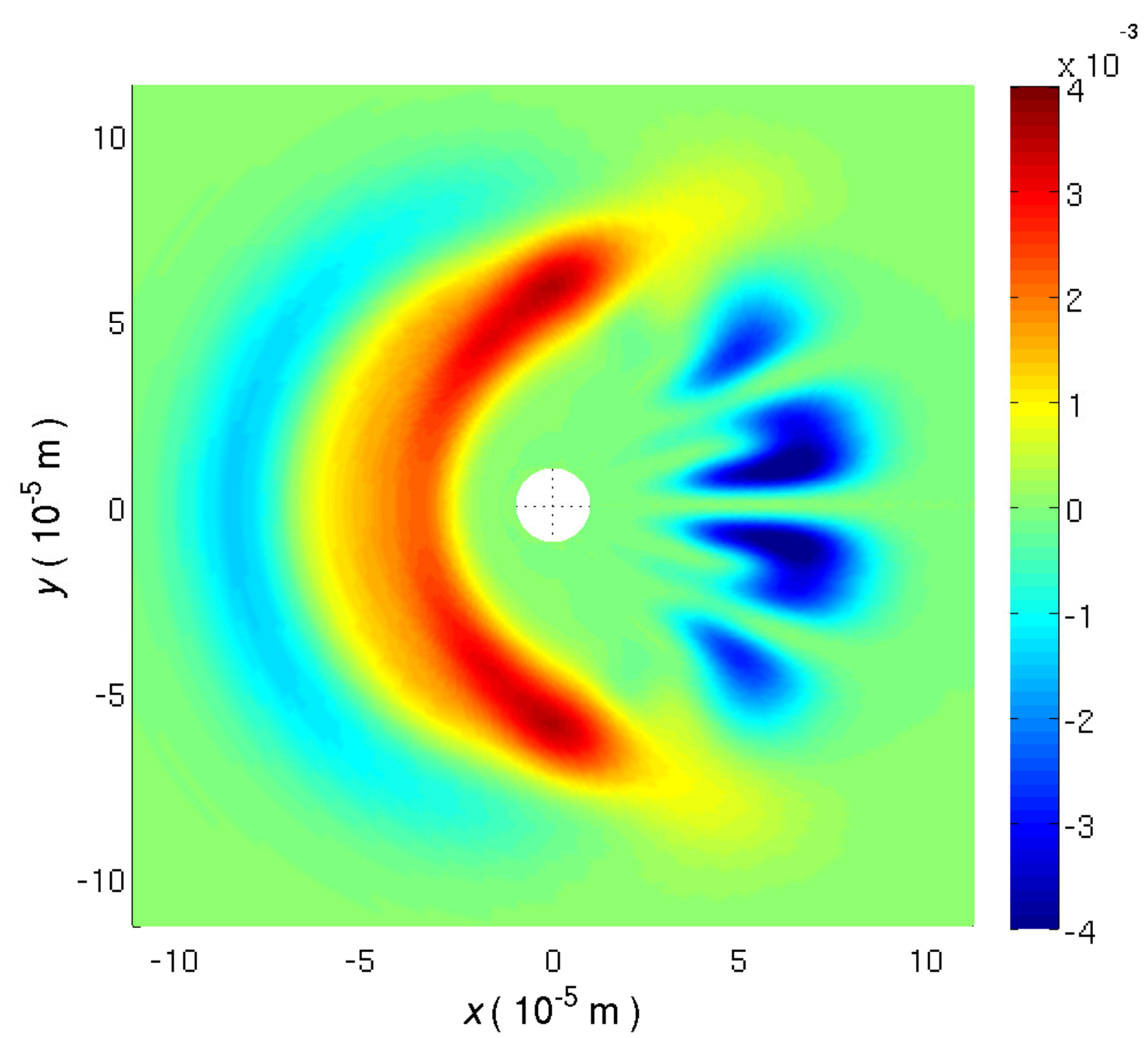}
\caption{{\bf Local spin density of AB-scattered unpolarized neutrons.} The incoming 
perpendicular velocity of the neutrons is $0.02$ m/s with the same wire radius and 
electrical current as in Fig \ref{fig:1}. Due to the attractive (repulsive) scalar potential 
felt by the local spin down (up) state, the measurable AB effect is fully dominated by 
the spin down state.}
\label{fig:3}
\end{figure*}

A significant simplification of the experimental setting is that unpolarized neutrons 
can be used, due to the spatial bifurcation caused by the energy splitting of the two 
local spin eigenstates. We demonstrate this effect in terms of the local spin density 
\begin{eqnarray}
\boldsymbol{\sigma} ({\bf r},s) & = &  
\frac{1}{2} \left( \left| \varphi_+ ({\bf r},s) \right|^2 \bra{\chi_+} \boldsymbol{\sigma} 
\ket{\chi_+} \right. 
\nonumber \\ 
 & & \left. + \left| \varphi_- ({\bf r},s) \right|^2 \bra{\chi_-} \boldsymbol{\sigma} 
\ket{\chi_-} \right) 
\nonumber \\ 
 & = & \frac{1}{2} \left( \left| \varphi_+ ({\bf r},s) \right|^2 - 
\left| \varphi_- ({\bf r},s) \right|^2 \right) {\bf e}_{\theta} . 
\end{eqnarray} 
The scalar part $\frac{1}{2} \left( \left| \varphi_+ ({\bf r},s) \right|^2 - \left| 
\varphi_- ({\bf r},s) \right|^2 \right)$ is shown Fig. \ref{fig:3} (see also Supplementary 
Movie 3). We note that only $\ket{\chi_{-} (\theta)}$ contributes significantly to the AB 
phase shift; the reason being the attractive nature of the adiabatic potential $-V_0 f(r)$ 
associated with this state for $r>1$.  

The low neutron velocities can be experimentally achieved by letting the neutron hit 
the wire at small angle. This can be realized for cold neutrons (speed a few 100 m/s) 
initially moving parallel to a horizontal wire. The bending of the neutron beam in the 
gravitational field induces a small velocity component $v_{xy}$ towards the wire. For 
fixed incoming speed, the relative velocity $v_{xy}$ can be controlled either by moving 
the wire in the vertical direction or by tilting it slightly upwards in the direction of 
motion. 

The incoming wave packet used in the numerical simulations have been chosen to clearly 
demonstrate the adiabatic AB effect on the spatial distribution of the scattered neutrons. 
However, to reproduce this wave packet dynamics in an actual experiment would be 
challenging as it would require a very high angular resolution and precise centering of 
the neutron beam. The difficulty is associated with the small size of the wave packet that 
would decrease the number of neutrons that would hit the wire in the desired manner. 
A less demanding setting would require the transversal width of the wave packet to be much 
larger than the width of the wire as this would allow for higher neutron flux. The AB-induced 
destructive interference effect in the forward direction would be intact for such a wave packet, 
as the AB effect is topological and therefore independent of shape and size of the neutron 
wave function. 

\section{Conclusions}
We have demonstrated an adiabatic AB effect for slow neutrons that scatter 
on a magnetic field produced by an electrical current through a very long wire. The 
mechanism of the effect is the Berry phase of the neutron spin restricted to the plane 
perpendicular to the wire. Therefore, the acquired phase shift is restricted to $\pi$, 
known to be the only possible non-trivial value of a planar spin. The $\pi$ phase shift 
causes destructive intereference in the forward direction, providing an unambigous 
signature of the adiabatic AB effect in a scattering setup. We have further demonstrated 
the nondispersive nature of the effect, which opens up for the possibility to observe 
the effect for higher neutron velocities in the adiabatic regime. 

\section*{Acknowledgments} 
E. S. acknowledges support from the Swedish Research Council through Grant No. D0413201. 
B. H. was supported by the National Research Foundation and the Ministry of Education 
(Singapore).

\end{document}